\documentclass{iopart}
\usepackage[dvips]{graphicx}

\begin{document}
\title[Electron affinity and ionization potential of carbon nanotubes]{\textit{Ab initio} calculations of electron affinity and ionization potential of carbon nanotubes}
\author{F Buonocore$^1$, F Trani$^2$, D Ninno$^2$, A Di Matteo$^3$, G Cantele$^2$ and G Iadonisi$^2$}
\address{$^1$ STMicroelectronics, Stradale Primosole 50, I-95121 Catania, Italy}
\address{$^2$ Coherentia CNR-INFM and Universit\`a di Napoli Federico II, Dipartimento di
Scienze Fisiche, Complesso Universitario Monte S. Angelo, Via Cintia, I-80126 Napoli, Italy.}
\address{$^3$ STMicroelectronics c/o IMAST P.le Enrico Fermi 1, Localit\`a Granatello Portici, I-80055 Napoli, Italy}
\begin{abstract}
By combining \textit{ab initio} all-electron localized orbital and pseudopotential plane-wave approaches we report on calculations of the electron affinity (EA) and the ionization potential (IP) of (5,5) and (7,0) single-wall carbon nanotubes. The role played by finite-size effects and nanotube termination has been analyzed by comparing several hydrogen-passivated and not passivated nanotube segments.
The dependence of EA and IP on both quantum confinement effect, due to the nanotube finite length, and charge accumulation on the edges, is studied in detail. Also, EA and IP are compared to the energies of the lowest unoccupied and highest occupied states, respectively, upon increasing the nanotube length. We report a slow convergence with respect to the number of atoms. The effect of nanotube packing in arrays on the electronic properties is eventually elucidated as a function of the intertube distance.
\end{abstract}
\ead{fabio.trani@unina.it}
\maketitle

\section{Introduction}
Interfacing carbon nanotubes either with a metal or with a semiconductor is a challenging step in the development of nanoscale electronic devices \cite{leonard06}. Apart from important aspects related to the formation of dipoles at the interface, it is well known that the electronic structure alignment is largely determined by the mismatch between the Fermi level of the substrate and the electron affinity/ionization potential of the adsorbed material. These are the driving concepts in the design of rectifiers, $p$-$n$ junctions and transistors \cite{leonard06}.

Most of the published theoretical papers discussing the properties of nanotubes are concentrated on \textit{ab initio} calculations of the work function (WF). Its dependence on the size, chirality and orientation has been studied with some detail \cite{bin,chun,su,agraval}. Bundles of nanotubes have also been considered \cite{su,reich}. Although the numerical results appear to suffer some spread due to the use of different WF definitions and different computational schemes, the general trend emerging from these papers is that, with the exception of small diameter nanotubes, the work function does not dramatically change with the tube chirality \cite{bin} and capping \cite{chun}. Surprisingly, the electron affinity (EA) and the ionization potential (IP), although related to the WF, have not received much attention, neither it has been discussed the interplay between these quantities and the presence of edge localized states. It is expected that these states may play an important role in interfacing a carbon nanotube with another material.

Field emission properties depend on the EA and IP as well. In particular, the huge aspect ratio (height to diameter) of carbon nanotubes makes them a very promising material for realizing low threshold voltage field emitters, such as lamps, X-ray tubes and flat panel displays. A rich literature has been flourishing in the last few years on this kind of applications. In this context, it is worth pointing out that although early studies reported field emission from samples where the carbon nanotubes were dispersed in the substrate \cite{wang98}, more recent papers report an excellent nanotube vertical alignment with homogeneous length and radius \cite{li99}. Moreover, the development of nanopatterning techniques for catalyst deposition opens the way to the fabrication of nanotube arrays with a predefined geometry and
a nanometre scale intertube distance.
As we shall see in the following, when the intertube distance is of the order of a few angstroms, the interactions between the nanotubes give rise to a band structure whose main features depend on the nanotube geometrical properties (chirality, either open or close edges and so on).

Motivated from the above considerations, in this paper we investigate the electronic properties of single-wall carbon nanotubes with methods based on the density functional theory. We have analyzed two classes of systems: the isolated nanotube and the corresponding periodic array. In the first case we put the emphasis on the dependence of EA and IP on both the nanotube geometry (either armchair or zig-zag) and length. In the second case we calculated the array band structure and the variations of EA and IP with the intertube distance.
We show the electronegativity as a function of length, in the case of isolated nanotubes, and the work function as a function of intertube distance, in the case of nanotube arrays.

\section{Computational details}
The \textit{ab initio} calculations have been performed using two different computational schemes. The first one is an all-electron method as implemented into the DMol$^3$ package (Accelrys Inc.) \cite{delley90,delley00} which makes use of a localized basis set. As such, the package is particularly useful in studying confined and isolated systems. The second scheme is based on a pseudopotential plane-wave method as implemented in the QUANTUM-ESPRESSO code \cite{espresso}. The latter code well suits the study of periodic systems such as the nanotube arrays. Our experience in using such mixed computational schemes is that both methods give results in good agreement, provided that i) the same exchange and correlation functional is used, and ii) the numerical convergence is carefully verified, with a large supercell for the plane-wave calculation and an accurate basis set for the localized orbital approach \cite{fest05}.

All calculations have been performed using the generalized gradient approximation (GGA) with the Perdew, Burke and Ernzerhof (PBE) correlation functional \cite{perdew96}. For the all-electron localized orbital calculations the electronic wave functions are expanded in atom-centred basis functions defined on a dense numerical grid. The chosen basis set was the Double Numerical plus polarization \cite{andzelm01}. This basis is composed of two numerical functions per valence orbital, supplemented by a polarization function, including a polarization $p$-function on the hydrogen atoms. The pseudopotential plane-wave calculations have been performed using Rabe-Rappe-Kaxiras-Joannopoulos (RRKJ) ultrasoft pseudopotentials \cite{rrkj}, a 26 Ry cut-off for the wave functions and a 156 Ry cut-off for the charge density. Since the nanotubes considered for the array calculations have a maximum relaxed length of 10.46 {\AA}, we have used a supercell whose size along the nanotube axis has been fixed to 35 {\AA}. We have checked that such supercell height is large enough to avoid spurious interactions between the array periodic replicas. We have also verified that a $2\times2$ Monkhorst-Pack $k$-point grid is a good choice in order to have converged total energies and band structures. For both the calculation schemes, the geometry optimization was done
relaxing all the atoms in the structure with a convergence threshold of 0.001 Ry/{\AA} on the interatomic forces.

\section{Results and discussion}

\subsection{Isolated Nanotubes}
As previously mentioned, an all-electron localized orbital approach has been
employed to address the electronic structure of isolated nanotubes. Several aspects have
been pointed out: nanotube geometry, by comparing zig-zag (7,0) and armchair (5,5)
nanotubes; nanotube length; edge termination, by considering both H-passivated
(H-pass) and not passivated (no-pass) edges.

\begin{figure}
 \begin{center}
\includegraphics[width=\textwidth]{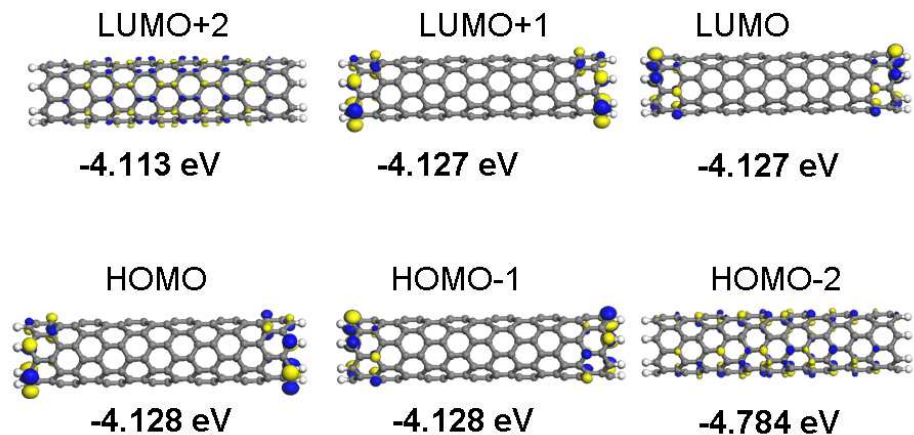}
\end{center}
 \caption{Molecular orbitals and energy levels of the H-pass (7,0) carbon nanotube. The isosurfaces correspond to 20\% of the maximum value. The optimized nanotube length is 26.33 {\AA}.}
\label{fig1}
\end{figure}

In figure \ref{fig1} we plot the three-dimensional contour plots of the squared wave functions
of the electronic orbitals of the H-pass (7,0) nanotube, with energies around the HOMO
(Highest Occupied Molecular Orbital) and the LUMO (Lowest Unoccupied Molecular Orbital).
In these calculations the nanotube is 26.33 {\AA} long.
It can be noted that four almost degenerate orbitals show up (HOMO-1, HOMO, LUMO, LUMO+1),
with edge-localized charge densities.
The H-pass (5,5) nanotube (not shown in figure), does not have edge localized orbitals near the Fermi level.
These results are consistent with those of H-pass graphene ribbons. Indeed, in \cite{nakada96} an analytic expression for the electronic wave functions of the edge states for graphene ribbons has been derived in the case of zigzag edges. It has been shown that these edge states have a topological nature and they were not predicted for armchair structures, in agreement with our findings. As one would expect, the pattern of the edge states completely changes when the passivating hydrogen atoms are removed. In this case the orbitals of both the (5,5) and (7,0) nanotubes exhibit edge localized states mainly due to the presence of dangling bonds. We have found that the no-pass (5,5) nanotube shows delocalized HOMO and LUMO orbitals, together with four almost degenerate edge localized orbitals, lying at 0.285 eV above the LUMO. The no-pass (7,0) has an even richer number of localized orbitals. In this case there is a loss of symmetry in the geometry, and the nanotube shows non degenerate LUMO and HOMO energy levels, localized on the open edges. We shall see in the following that the complex interplay between edge localized and delocalized orbitals has some influence on both the EA and IP and, in the case of a nanotube array, on the band structure.

\begin{figure}
\begin{center}
\includegraphics[width=\textwidth]{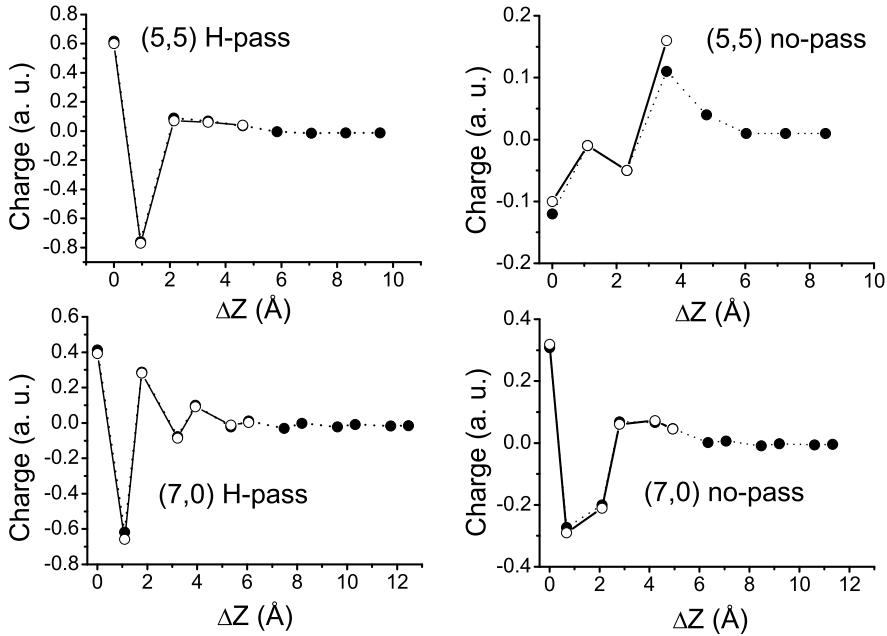}
\end{center}
\caption{Total Mulliken charge (in atomic units) calculated on planes perpendicular to the nanotube axis
as a function of the distance $\Delta z$ from the edge \cite{note}.
For each nanotube, the calculation has been done for two different lengths
(full and open circles). At the end of the geometrical optimization, the lengths (expressed in {\AA})
are as it follows: H-pass (5,5), 20.31 and 10.46; no-pass (5,5), 18.20 and 8.36; H-pass (7,0), 25.99 and 13.54;
no-pass (7,0), 24.31 and 11.52.}
\label{fig2}
\end{figure}

In figure \ref{fig2} we plot the total Mulliken charges computed on atomic planes perpendicular to the nanotube axis for (5,5) and (7,0) nanotubes with different lengths.
Because of symmetry, the plot is limited to half the distance from the nanotube edges. The Mulliken charges near the nanotube edge have only a very weak dependence on the nanotube length, whereas significant differences due to the nanotube geometry and edge terminations are observed. The H-pass (5,5) has a dipole near the edge reflecting the ability of the hydrogen atom to donate its electron. Because of the armchair shape, the edge carbon atoms tend to form a double bond. In the no-pass (5,5) this dipole is reversed due to some electronic charge transfer from inside the nanotube to the edge. And the armchair carbon atoms give rise to triple bonds. A good indication supporting this interpretation emerges from the calculation of the C-C bond lengths at the nanotube edge. For the H-pass (5,5) nanotube, the 1.37 {\AA} long C-C distance can be compared with the 1.33 {\AA} long double bond of C$_2$H$_4$. For the no-pass (5,5) nanotube, the 1.24 {\AA} long bond length can well be compared with the 1.20 {\AA} long triple bond in C$_2$H$_2$. Instead, in the case of H-pass and no-pass (7,0) nanotubes, the C-C bond length is in the range 1.42$\div$1.44 {\AA} (a value close to that one in graphene). Figure \ref{fig2} shows that the (7,0) nanotube has a different behaviour, since there is not dipole flipping on the edge. This is due to the zig-zag shape of the edge which tends to preserve the graphene bonding pattern. By the way, it should be noted that the removal of the hydrogen atoms from the edge does induce charge redistribution in the first couple of atomic planes.

\begin{figure}
\begin{center}
\includegraphics[width=0.5\textwidth]{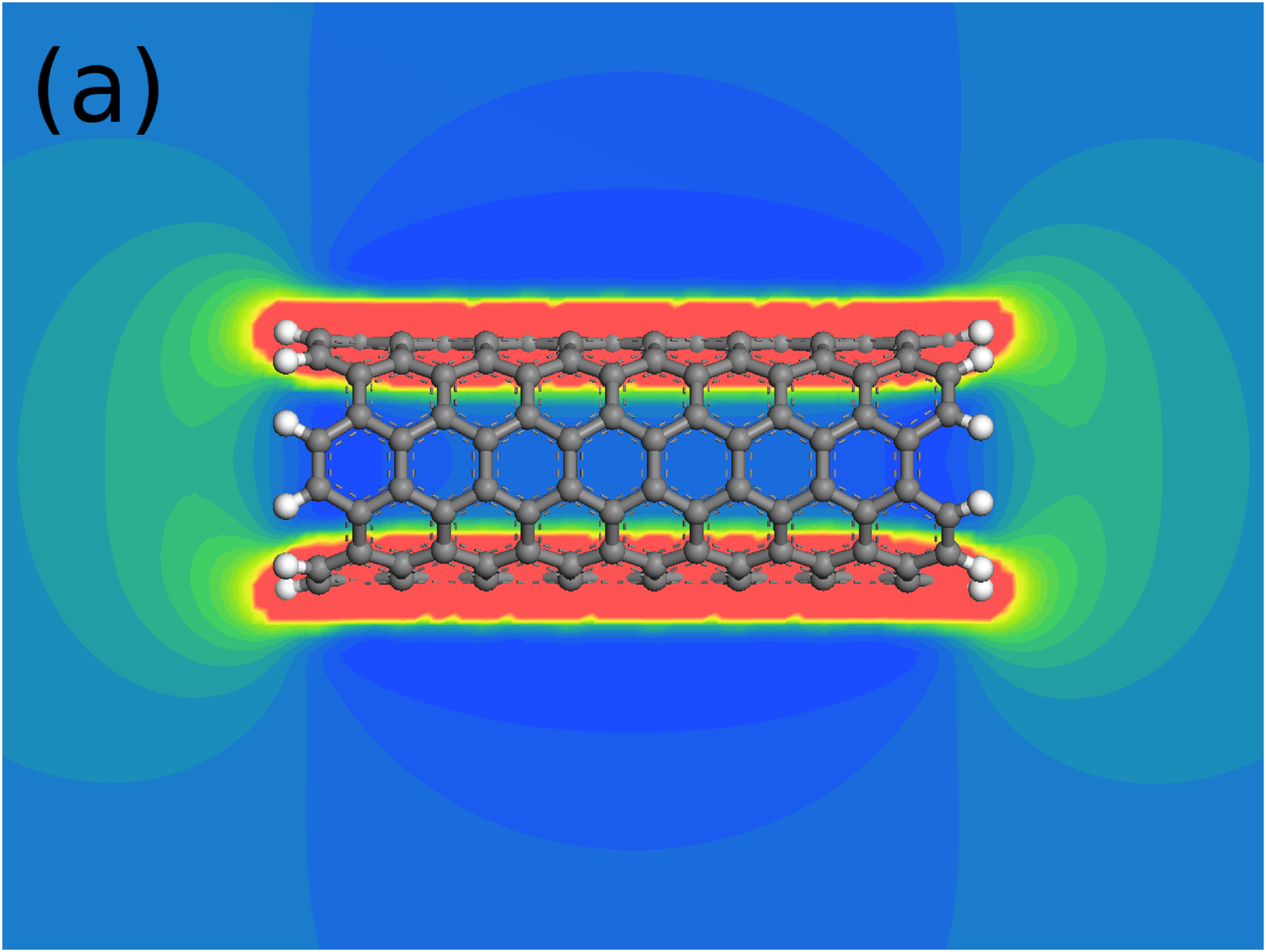}\includegraphics[width=0.5\textwidth]{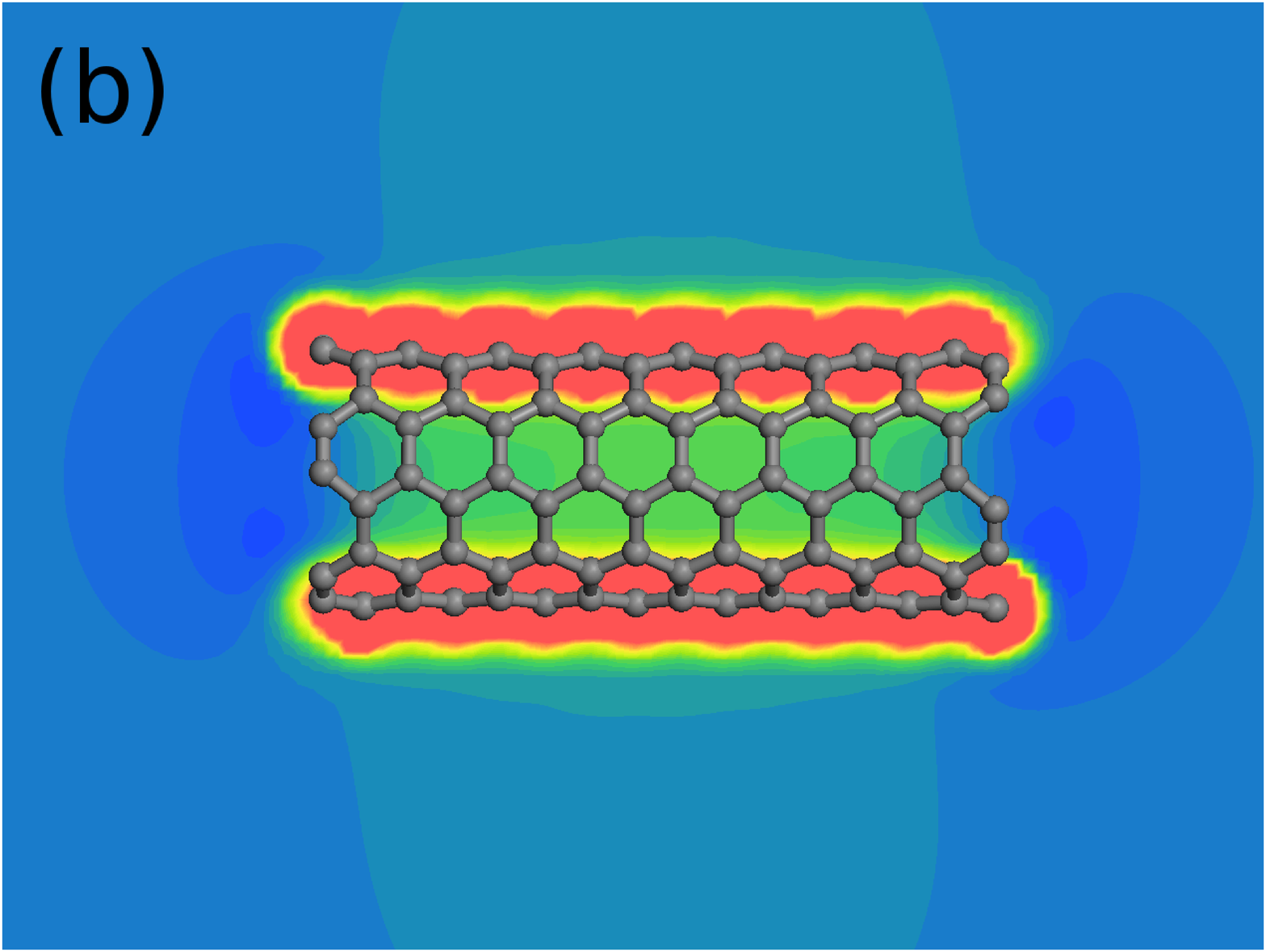}
\end{center}
\caption{Planar contour plots of the all-electron electrostatic potential energy of a (5,5) carbon nanotube with (panel a) and without (panel b) hydrogen atoms on the edge. The potential energy is the lowest (highest) in the red (blue) regions. The optimized lengths of these structures are 20.31 {\AA} (panel a) and 18.20  {\AA} (panel b). }
\label{fig3}
\end{figure}

In order to give a more detailed account of the effects induced by the charge distribution
highlighted by the Mulliken analysis, we show in figure \ref{fig3}
a two-dimensional contour plot of the all-electron electrostatic potential energy calculated for a (5,5)
nanotube, in both H-pass and no-pass configurations.
It comes out that the edge passivation strongly affects the nanotube electronic properties.
The potential energy feels a sharp variation close to the nanotube cage, when we move along a direction
orthogonal to the symmetry axis, for the H-pass nanotube (panel a). A similar sharp variation is seen 
close to the edges of the non passivated nanotube (panel b), when moving along the direction
parallel to the axis.
Another interesting feature is the potential distribution inside the nanotube cage. The hydrogen passivation lowers the potential energy on the edges. Instead, the C-C bonding occurring when hydrogen atoms are removed lowers the potential inside the cage. These findings are fully consistent with the charge distributions of figure \ref{fig2}.

Let us start the discussion of the numerical results with the work function (WF), for which several theoretical and experimental data are available. The WF of a periodic system is calculated as  WF = E$_{vac}$- (E$_{LUMO}$+E$_{HOMO}$)/2 (the vacuum energy E$_{vac}$ is defined as the electrostatic potential energy in the vacuum, far away from the system). This definition has been applied to the calculation of the WF of infinite isolated nanotubes, as well as to the case of two-dimensional finite nanotube arrays, discussed in the second part of the paper. The infinite (5,5) nanotube is a zero gap metal for which we have obtained an all-electron WF of 4.37 eV. The plane-wave calculation gives 4.28 eV. For the infinite (7,0) nanotube we have an all-electron WF of 4.82 eV to be compared with the plane-wave result of 4.75 eV. Although in both cases the two methods are in good agreement, it should be mentioned that the WF strongly depends on the exchange and correlation functional. It is likely that this is the reason why the present values of WF are smaller than those shown in \cite{bin}. In any case, the comparison with the many experimental available data is acceptable. TEM measurements on multiwall nanotubes give 4.6-4.8 eV  \cite{gao}, photoelectron emission gives 4.95 eV and 5.05 eV for multi- and single-wall  \cite{shiraishi}, thermionic emission for multi-wall gives 4.54-4.64 eV  \cite{peng}, UPS measurements on single-wall give 4.8 eV \cite{suzuki}.

In a periodic system, the electron affinity is calculated as EA=E$_{vac}$-E$_{LUMO}$ and the ionization potential as IP=E$_{vac}$-E$_{HOMO}$. However, a different definition holds for finite systems. In this case, the electron affinity is defined as EA=E(N)-E(N+1) where E(N) and E(N+1) are the total ground-state energies in the neutral (N) and single charged (N+1) configurations. The ionization potential is similarly defined as IP=E(N-1)-E(N). A great deal of theoretical work has been devoted to understand how the two definitions are connected \cite{janak78, perdew82, perdew97, filippetti98, zhang, zhan}. It is a common use to apply the second approach to isolated systems, such as molecules and nanocrystals \cite{zhan,chely}; in the case of extended systems, instead, the IP and EA are calculated from the first approach \cite{bernholc}. Although both the definitions should converge to the same value in the limit in which a finite system tends to an infinite one \cite{filippetti98}, a numerical check of this convergence is almost impossible due to the huge computational cost of this operation. Moreover, there are cases where even the second definition could fail giving wrong results for the electron affinity \cite{filippetti98}.
Once the EA and IP of a finite nanotube have been calculated, a measure of the nanotube reactivity may be inferred from the Mulliken electronegativity, defined as $\chi$=(EA+IP)/2 \cite{kohn}. An interesting aspect of this definition is that it is identical to the work function of an infinite nanotube assuming, for the case of a semiconducting tube, a Fermi level sitting at midgap.

\begin{figure}
\begin{center}
\includegraphics[width=0.8\textwidth]{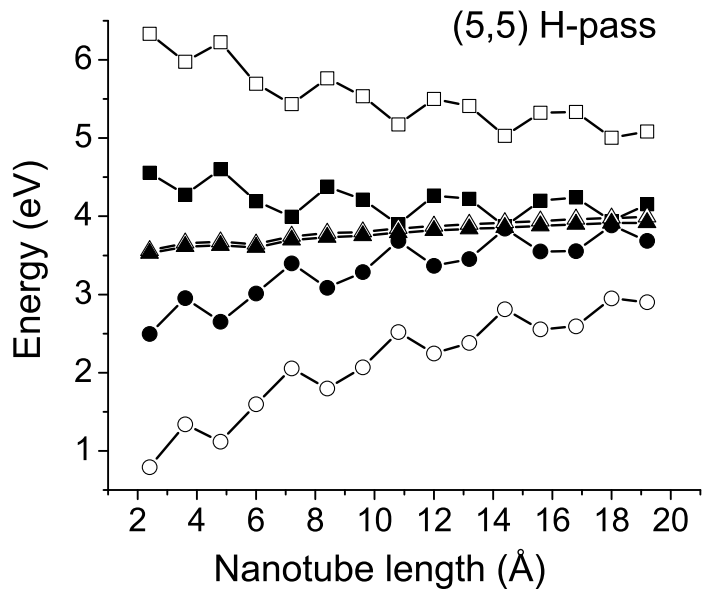}
\end{center}
\caption{Ionization potential (squares), electron affinity (circles) and electronegativity (triangles) of a H-pass (5,5) nanotube as a function of its length. The results represented by open symbols are obtained from total energies whereas those indicated with full symbols are derived from the HOMO and LUMO energy levels.}
\label{fig4}
\end{figure}

In a finite nanotube, quantum confinement may induce a strong dependence of the electronic properties on the nanotube length. In order to give a clear insight on these effects, we have calculated the EA and IP for a number of tubes with increasing length. The results are shown in figure \ref{fig4} for a (5,5) nanotube with the edges passivated with hydrogen atoms.
In this figure both EA (circles) and IP (squares) have been calculated starting from either
total energies (open symbols) or LUMO/HOMO energy levels (full symbols).
The first observation to be made is that both EA and IP, as well as the gap, exhibit a regular oscillation on increasing the nanotube length. Nonetheless, the quantum confinement effect is evidenced
by the tendency to reduce the IP-EA difference for long nanotubes. Moreover,
the oscillation pattern does not change when EA and IP are calculated with different methods. From the point of view of the HOMO-LUMO gap,
the cause of these oscillations has been discussed in \cite{lu04}, where it is shown,
within a tight binding approach, that the gap of finite armchair nanotubes vanishes
every 3 sections (see \cite{lu04} for the definition of a section).
Actually, from our calculations it is found that the energy gap does
not completely vanish, giving indication that the interactions go well beyond the first
few nearest-neighbours. An even more interesting finding is that the electronegativity  is not influenced by these strong oscillations and it is nearly independent of the definitions of both EA and IP. Over the explored range of nanotube lengths, $\chi$ has an overall variation of about 0.5-0.6 eV.
A similar dependence on the nanotube length of both EA and IP
is found for the no-pass (5,5). However, because of the flipping of the nanotube edge dipoles shown in figure
\ref{fig2}, EA, IP and $\chi$ are raised in energy. For instance, for the longest
nanotube of figure \ref{fig4}, we have found a 0.6 eV rise of the EA and IP with respect to the H-pass nanotube.

\begin{figure}
\begin{center}
\includegraphics[width=0.8\textwidth]{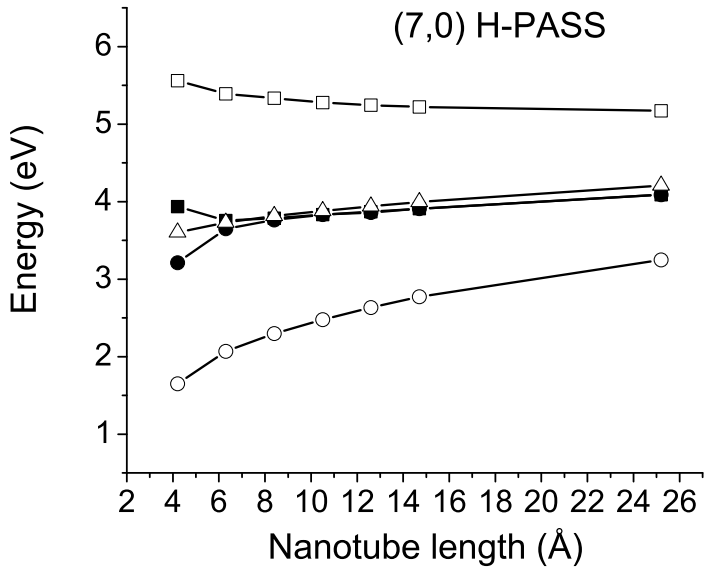}
\end{center}
\caption{Ionization potential (squares), electron affinity (circles) and electronegativity (triangles) of no-pass (7,0) nanotube as a function of the nanotube length. The results represented by open symbols are obtained from total energies whereas those indicated with full symbols are derived from the HOMO and LUMO energy levels.}
\label{fig5}
\end{figure}

In figure \ref{fig5} we show the IP and EA for a (7,0) H-pass nanotube. 
In such a configuration, as shown in figure \ref{fig1}, the HOMO and LUMO levels are localized on the edge. And therefore, with the exception of very small tubes, they are located within a narrow energy window and so are the IP, EA and $\chi$.
A different result is obtained, with well separated values, when EA, IP and $\chi$ are calculated from the total energies.
However, despite the deviations shown in figure \ref{fig5}, the electronegativity seems, again, to be independent of the definitions used for EA and IP.

We close this section with a final comment on the results of Figs.
\ref{fig4} and \ref{fig5}. Looking at the differences between HOMO and IP on one side and LUMO and EA on the other, it is evident that the many-body corrections to GGA are significant in the considered size range.
For the (5,5) H-pass of figure \ref{fig4} the self energy correction
defined as the difference between the quasi-particle gap (EA-IP from total energies) and the HOMO-LUMO gap
ranges between 3.4 and 1.7 eV going from the shortest to the longest nanotubes. Similarly, for the (7,0) H-pass of figure \ref{fig5} the variation ranges from 3.2 to 1.9 eV.
\begin{figure}
\begin{center}
\includegraphics[width=\textwidth]{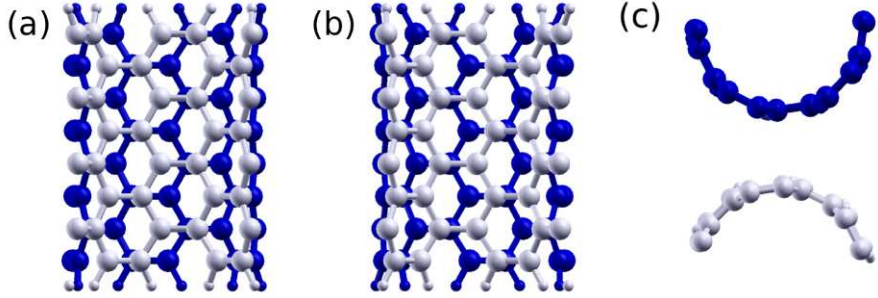}
\end{center}
\caption{Views along a) $x$, b) $y$ and c) $z$ of two adjacent nanotubes constituting the array of H-pass (5,5) nanotubes. Dark (blue) and light (white) spheres indicate back and front atoms, respectively. Hydrogen atoms are represented by the smaller spheres.}
\label{fig6}
\end{figure}
\subsection{Finite Nanotube Arrays}
We have studied (5,5) finite-size nanotube arrays using the QUANTUM-ESPRESSO package \cite{espresso}.
The nanotubes are organized in such a way to compose a 2D square lattice. At the end of the geometrical optimization, in which all the atoms in the unit cell are allowed to relax, the nanotubes examined are about 8.4 \AA{} and 10.5 \AA{} long, in the not passivated and hydrogen passivated configuration, respectively.
When constructing the array, the first decision one has to make is on how to take the relative orientation of the nanotubes. This is an interesting point whose assessment requires an accurate description of the long-range (van der Waals) interaction between nanotubes. In a recent paper \cite{carlson} the modeling of the intertube interaction has been studied within a tight binding scheme showing that two adjacent and parallel nanotubes have a minimum energy when are rotated in such a way to have a stacking similar to that one of graphite. Unfortunately, such a stacking could be treated within our DFT scheme only using
very large supercells making the calculation unpractical. Nevertheless, we have done a series of total energy calculations by rotating around its axis and relaxing just the nanotube in the unit cell. Although the long range intertube interactions are not well represented in the present GGA calculations, we have found a minimum in the total energy when the nanotubes have a stacking very similar to that of \cite{carlson}. The result is shown in figure \ref{fig6} through three views of two nanotubes belonging to two adjacent unit cells of the H-pass (5,5) array. The similarity to the graphite stacking is evident.

\begin{figure}
\begin{center}
\includegraphics[width=0.7\textwidth]{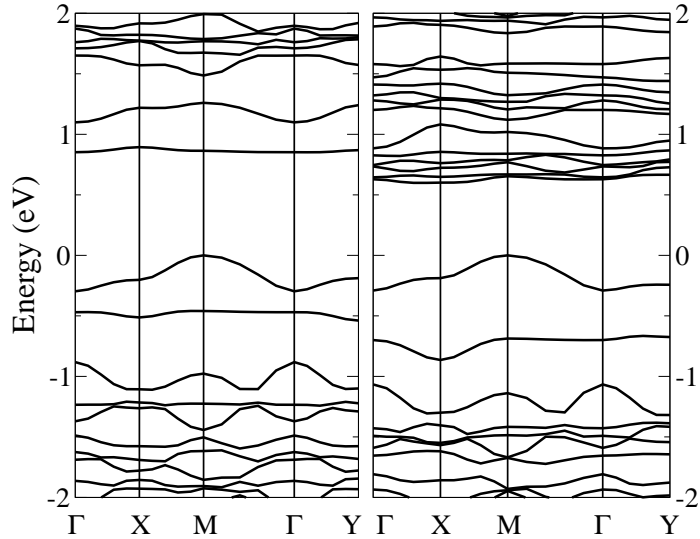}
\end{center}
\caption{Band structures of H-pass (left) and no-pass (right) (5,5) nanotube arrays. The lattice constant is $10$ {\AA} long. }
\label{fig7}
\end{figure}

In figure \ref{fig7} the square array band structures of H-pass (left panel) and no-pass (right panel) (5,5) nanotubes are shown. The array lattice constant has been fixed to $a=10$ {\AA},
corresponding to a minimum wall-wall distance of 3.2 {\AA}. The usual notation of the reciprocal square lattice irreducible wedge has been used with the top valence band chosen as the zero energy. An interesting result coming from figure \ref{fig7} is that the top valence band depends very little on the edge passivation, with the highest occupied state at the M point. The bottom conduction band is dispersionless and, contrary to the valence band, it has a significant dependence on whether or not the nanotube edges are terminated with hydrogen atoms.
In particular, the number of conduction bands near the energy gap increases in the no-pass nanotube for the presence of dangling bonds. For the H-pass nanotube array, the first conduction band is well isolated, with a small dispersion of 0.04 eV, and a minimum close to the $\Gamma$ point. The band gap is 0.85 eV. For the no-pass nanotube array, there is a group of 5 quasi-degenerate levels at the bottom of the conduction band, all together having a dispersion of about 0.22 eV, with a minimum close to the X point. The band gap in this case is 0.6 eV, thus smaller than in the H-pass case. So, the removal of the hydrogen atoms from the edges leads to i) an increase of the first conduction band dispersion and quasi-degeneration, ii) a decrease of the band gap, iii) a change in the position of the bottom of the conduction band.

\begin{figure}
\begin{center}
\includegraphics[width=0.7\textwidth]{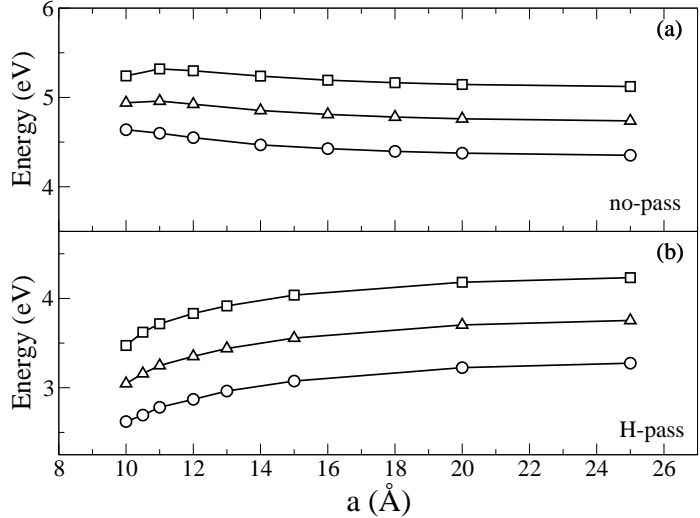}
\end{center}
\caption{Ionization potential (squares), electron affinity (circles) and work function (triangles) for no-passivated (panel a) and H-passivated (panel b) (5,5) nanotube array, as a function of the array lattice parameter $a$. Lines are guides for the eyes.}
\label{fig8}
\end{figure}

As pointed above, the EA, IP and WF of nanotube arrays are calculated using the definition usually applied to periodic systems, based on the LUMO and HOMO energies. In figure \ref{fig8} the EA, IP and WF of no-pass (panel a) and H-pass (panel b)
nanotube arrays are shown as a function of the array lattice parameter.
Interesting differences between the no-pass and the H-pass nanotube arrays arise.
The trend in figure \ref{fig8} can be easily discussed in terms of nanotube edge dipoles.
The charge accumulation highlighted in figure \ref{fig2}, gives rise to edge dipoles that,
with their orientation, control the work function. When an array of nanotubes is formed, the surface density of dipole increases on reducing the array lattice spacing. Therefore the work function can either increase or decrease on reducing the lattice spacing, according to the dipole orientation.

\section{Conclusion}

In this paper H-passivated and not passivated (5,5) and (7,0) finite-size carbon nanotubes have been studied using \textit{ab initio} calculations. It emerges that the EA and IP in finite carbon nanotubes are controlled by two concurrent effects. The first one is similar to a quantum confinement effect in that it gives a variation of both EA and IP with the nanotube length; the second is a purely electrostatic effect due to the formation of edge dipoles.
We have seen that the electronegativity, a quantity that may be related to the work function
of the extended nanotube, can be calculated either from the HOMO and LUMO energy levels, or from the total energy. This is due to the circumstance that the self-energy correction to the HOMO and LUMO energies are of similar amplitude.
In the case of a nanotube array, the third element that comes into play is the array density through which the number of dipoles per surface area may be varied. Both EA and IP can either increase or decrease (with respect to the isolated nanotube) depending on the dipoles density and orientation.
At least in principle, with the control at the nanoscale on both the nanotube length and array density
would allow to tune the electron affinity and the ionization potential. As a final remark, we would like to mention that in a recent paper \cite{suzuki04} the work function of individual single-wall carbon nanotubes has been measured with photoemission microscopy. By analyzing the data coming from a set of nanotubes, the authors have been able to conclude that most of them have work functions ranging within a 0.6 eV window.
Although it may be a fortuitous coincidence, the set of calculations presented in this work does give an overall work function variation in the range 0.5-0.6 eV.

\ack
This work has been realized within the GINT collaboration (Gruppo INFN per le NanoTecnologie).
Financial support from PRIN-MIUR 2005 project is acknowledged.
Computational resources have been provided by CINECA-``Progetti Supercalcolo 2007.''
F. Trani acknowledges financial support from the S.Co.P.E. project.

\section*{References}


\begin{thebibliography}{10}
\bibitem{note} It should be noticed that atoms belonging to the same plane
in the infinite nanotube, slightly relax in presence of edges. The amount
of relaxation depends on the edge passivation. In this cases we assign to each
plane a $z$ coordinate obtained by averaging on those atoms.
\end{thebibliography}
\providecommand{\newblock}{}

\end{document}